\documentclass{emulateapj}

\newcommand{\beq}{\begin{equation}}
\newcommand{\beqa}{\begin{eqnarray}}
\newcommand{\eeq}{\end{equation}}
\newcommand{\eeqa}{\end{eqnarray}}

\shorttitle{Upper bound on the cosmic abundances of the negative-mass
 object and the Ellis wormhole}
\shortauthors{R. Takahashi \& H. Asada}

\begin{document}

\title{Observational upper bound on the cosmic abundances of negative-mass
 compact objects and Ellis wormholes from the Sloan Digital 
 Sky Survey quasar lens search}

\author{Ryuichi Takahashi \& Hideki Asada}
\affil{Faculty of Science and Technology, 
Hirosaki University, Hirosaki 036-8561, Japan\\
}

\begin{abstract}
The latest result in the Sloan Digital Sky Survey 
Quasar Lens Search (SQLS) has set the first cosmological constraints 
on negative-mass compact objects and Ellis wormholes. 
There are no multiple images lensed by the above two 
exotic objects for 
$\sim 50000$ distant quasars in the SQLS data.
Therefore, an 
upper bound is put on the cosmic abundances of these lenses. 
The number density of negative mass compact objects is
 $n<10^{-8} (10^{-4}) h^3 {\rm Mpc}^{-3}$ at the mass scale
 $|M| > 10^{15} (10^{12}) M_\odot$, which corresponds to
 the cosmological density parameter $|\Omega| < 10^{-4}$ at the
 galaxy and cluster mass range $|M|=10^{12-15}M_\odot$. 
The number density of the Ellis wormhole is $n<10^{-4} h^3
 {\rm Mpc}^{-3}$ 
for a range of the throat radius $a = 10-10^{4}$pc, 
which is much smaller than the Einstein ring radius. 
\end{abstract}



\section{Introduction}

In theoretical physics, negative mass is a hypothetical concept 
of matter whose mass is of opposite sign to the mass of normal matter. 
Negative mass could thus generate 
the {\it repulsive} gravitational force.  
Although possible negative mass ideas 
have been often discussed since 
the 19th century, 
there has been no evidence for them 
(Bondi 1957; Jammer 1961, 1999).
Even if 
its gravitational mass is negative, its inertial mass can be positive
 or negative (e.g. Jammer 1961).
If the inertial mass is positive, the negative mass repels the ordinary
 matter (positive-mass objects) and 
hence it is likely to escape 
from the Milky Way.
The negative masses attract each other 
to form a {\it negative} massive clump.
Such clumps might reside 
in cosmological voids (e.g. Piran 1997).
If the inertial mass is negative, 
on the other hand, 
the negative mass 
acts gravitationally as 
the ordinary matter. 
In this case, the negative masses 
could thus reside 
in the Galactic halo.

The gravitational lensing by the negative mass is the same as 
 that by the positive mass, but its deflection angle has the opposite sign. 
Several authors have 
suggested that the negative masses could be detected
 in the Galactic microlensing (Cramer et al. 1995; Safonova et al. 2001a).
Torres et al. (1998) assumed that the lensing of the distant
 active galactic nuclei by the 
hypothetical negative mass could be 
detected as the
 gamma ray burst and they provided the constraint on its mass density 
 as $|\rho| \lesssim 10^{-36} {\rm g cm}^{-3}$
 (corresponding to the cosmological density parameter $|\Omega|
 \lesssim 10^{-7}$) around the mass scale of $|M| \sim 0.1 M_\odot$.

The wormhole is a hypothetical object connecting distant regions 
of the universe, like a space-time tunnel.
Ellis found a wormhole solution of the Einstein equation 
in general relativity by introducing a massless scalar field (Ellis 1973). 
Later, Morris \& Thorne (1988) and Morris, Thorne \& Yurtsever (1988)
 studied this solution as the traversable wormhole.
The energy condition would be violated in order to create and maintain the
 wormhole (Visser 1995). 
Dark energy that could violate the energy condition 
is introduced to explain the observed accelerated expansion of the universe. 
The Ellis wormhole is massless and does not 
gravitationally interact with 
ordinary 
matter at remote distance.  
Hence, it makes no contribution to 
the cosmic mass density, 
even if it lives in our universe. 
However, it can deflect the light path.
Recently, 
Abe (2010) has 
suggested that the wormhole at the throat
 radius of $100-10^7$km could be 
constrained (or detected) by 
using the Galactic microlensing.
Although its abundance is quite unknown, 
some authors speculated that it 
is as abundant as stars in the universe (Krasnikov 2000; Abe 2010). 
More recently, Yoo et al. (2013) gave the rough upper bound of its number
 density as $n \lesssim 10^{-9} {\rm AU}^{-3}$ for the throat radius
 $a \sim 1$cm from the femtolensing of distant gamma ray bursts
 (Barnacka et al. 2012).

Although there are a lot of theoretical works concerning 
negative-mass objects and Ellis wormholes 
(e.g., Morris \& Thorne 1988; Cramer et al. 1995; Visser 1995),
observational studies have been very rare, mainly because 
no matter accretion occurs 
owing to the repulsive force by the negative mass and the Ellis wormhole, 
and it is thus unlikely to directly see them as luminous objects.
Hence, it has recently attracted interests 
to study the gravitational lensing as an observational tool to 
probe such exotic dark objects (Kitamura et al. 2013; Tsukamoto \& Harada 
 2013).

When a light ray from a distant quasar passes near the above lens 
objects (the negative mass and the Ellis wormhole), 
multiple images of the quasar are formed without any normal lens object. 
The absence of such multiple images can 
limit 
the cosmological abundances of these lens objects.
The purpose of this letter is to place a first upper bound 
on the cosmic abundances of such exotic objects 
using the latest gravitational lensing survey. 
The SQLS has the current largest quasar lens sample from
 the SDSS II Data Release 7 (York et al. 2000).
There are $50,836$ quasars in the redshift range of $z=0.6-2.2$
 with the apparent magnitude brighter than $i=19.1$.
The SQLS searched the lens systems in the image angular separation of 
 $1^{\prime \prime} - 20^{\prime \prime}$ and
 found $19$ lensed quasars (Oguri et al. 2006, 2008, 2012; Inada et al. 2012).
Note that this is currently the largest homogeneous sample for all the
 wavelengths of light.
However, there is no lensed image candidate formed by unseen lens
 objects\footnote{This paper assumes that the negative masses are unseen. 
 However, they might form stars (or galaxies) to emit radiation like the
 usual matter. In this case, it would be difficult to distinguish the
 negative mass object from the positive one if these spectral energy
 distributions are the same.
 However the spectra of such negative-mass stars are quite unknown.}
 such as the cosmic string, the black hole, and the dark halo
 (Oguri \& Kayo, private communications).
Hence, the absence of the unusual lensed quasar gives a strong constraint on
 the abundances of such exotic lens objects.

Throughout this paper, we employ the units of $G=c=1$ and 
use the cosmological parameters of
 the Hubble constant $h=0.7$, the matter density $\Omega_{\rm m}=0.28$
 and the cosmological constant $\Omega_\Lambda=0.72$,
 which is in concordance with 
the latest WMAP 9yr result (Hinshaw et al. 2012).

\section{Gravitational lensing by negative-mass compact objects and 
Ellis wormholes} 

\subsection{Negative-mass compact object}
The lensing by the negative-mass compact object 
could be described 
by the Schwarzschild lens 
with its negative mass $M<0$. 
The lensing properties such as image positions and magnifications were
theoretically 
studied by Cramer et al. (1995) and Safonova et al. (2001a,b).
The image angular position $\theta$ and the source position $\beta$ are
 related via the lens equation,
\beq
  \beta=\theta + \frac{\theta_{\rm E}^2}{\theta},
\eeq
with the Einstein angular radius,
\beq
  \theta_{\rm E}= \left( {4  \left| M \right|} \frac{D_{LS}}{D_L D_S}
 \right)^{1/2},
\label{eins_rad_nega}
\eeq
where $D_L$ and $D_S$ are the angular-diameter distances to the lens and
 the source, $D_{LS}$ is the distance from the lens to the source.  
The number of images depends on the source position:
 no image forms for $\beta<2 \theta_{\rm E}$, one image forms for
 $\beta=2 \theta_{\rm E}$, and the two images form for
 $\beta > 2 \theta_{\rm E}$ at the positions
 $\theta_{\pm}= (\beta \pm \sqrt{\beta^2-4 \theta_{\rm E}^2} )/2$.
The image angular separation is $\Delta \theta = |\theta_+-\theta_-|$.
The magnification of each image is
 $\mu_{\pm}=|1-\theta_{\rm E}^4/\theta_\pm^4|$.

We obtain a relation between 
the typical lens mass 
and the image angular separation 
from Eq.(\ref{eins_rad_nega}) as,
 $|M| \simeq 1 \times 10^{11} h^{-1} M_\odot$
 $(\theta_{\rm E}/1^{\prime \prime})^2$ $[(D_L D_S/D_{LS})/1h^{-1}{\rm Gpc}]$.
As a result, the sensitive mass range is $|M|=10^{11-14} M_\odot$
 for the image separation of $1^{\prime \prime}-20^{\prime \prime}$.

\subsection{Ellis wormhole lens}

The Ellis
wormhole is known to be a massless wormhole, which means that the
asymptotic mass at infinity is zero. 
However, this wormhole deflects light by gravitational 
lensing (Cl\'ement 1984; Chetouani \& Cl\'ement 1984; Perlick 2004;
 Nandi et al. 2006; Dey \& Sen 2008; Nakajima \& Asada 2012;
 Gibbons \& Vyska 2012) because of its local curvature of the spacetime. 
The Ellis wormhole is characterized by 
one parameter as its throat radius $a$, and
 the line element is $ds^2 = dt^2 - dr^2 - (r^2 + a^2)
 (d\theta^2 + \sin^2\theta d\phi^2)$.
For Ellis wormholes, the lens equation in the weak field approximation 
 becomes,
\begin{equation}
 \beta = \theta - \theta_{\rm E}^3 \frac{\theta}{\left| \theta \right|^3},
\end{equation}
with the Einstein angular radius,
\beq
  \theta_{\rm E}=\left( \frac{\pi a^2}{4} \frac{D_{LS}}{D_L^2 D_S}
    \right)^{1/3}.
\label{eins_rad_ellis}
\eeq
The two images form irrespective of the source position
 at the angular positions,
 $x_\pm^{-1} =$ $\pm \sqrt[3]{1/2 +\sqrt{1/4 \pm y^3/27}}$
 $-\sqrt[3]{\mp  1/2 \pm \sqrt{1/4 \pm y^3/27}}$,
where $x_\pm=\theta_\pm/\theta_{\rm E}$ and $y=\beta/\theta_{\rm E}$
 (Toki et al. 2011; Nakajima \& Asada 2012).
The magnification for each image is
 $\mu_\pm=|(1 \mp x_\pm^{-3}) (2 x_\pm^{-3} \pm 1)|^{-1}$.

The typical throat radius 
for a given $\theta_{\rm E}$ 
is estimated as, $a \simeq 10h^{-1} {\rm pc}$
 $(\theta_{\rm E}/1^{\prime \prime})^{3/2}$
 $[(D_L^2 D_S/D_{LS})/(1h^{-1} {\rm Gpc})^2]^{1/2}$, from
 Eq.(\ref{eins_rad_ellis}).
As a result, the sensitive throat radius is $a=10-100$pc 
 from the image separation of $1^{\prime \prime}-20^{\prime \prime}$.
Note that the relevant throat radius is much smaller 
than the corresponding Einstein ring radius 
$R_{\rm E} = \theta_{\rm E} \times D_{\rm L} \sim 
10 {\rm kpc} (\theta_{\rm E}/1^{\prime \prime}) 
(D_L/1h^{-1} {\rm Gpc})$. 
In practice, the range of the radius is slightly wider, 
since the lensing effects are dependent also on the distance ratios.

\section{Strong Lensing Probability}

\begin{figure}
\epsscale{1.12}
\plotone{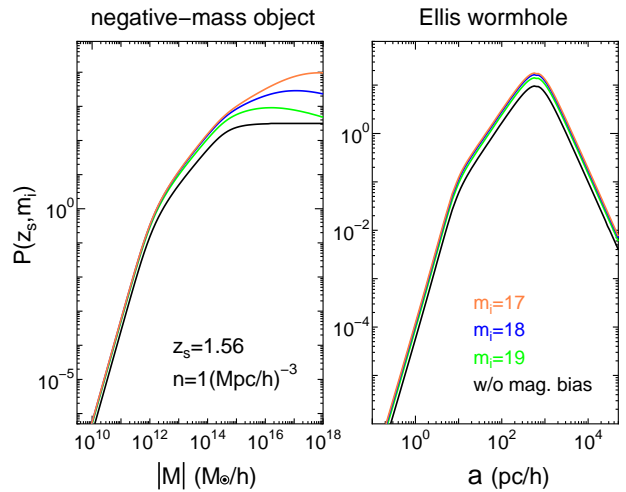}
\caption{
Multiple lensing probability that a quasar at the redshift $z_s$ is lensed
 by a 
foreground lens object. 
Left: Negative-mass objects. 
Right: Ellis wormholes.  
The horizontal axis is the absolute value of mass $|M|$ (left panel) and
 the throat radius $a$ (right panel). 
The source redshift is set to be $z_s=1.56$ 
that is the mean redshift 
 of statistical sample.
We consider three cases for 
the apparent magnitude of the observed quasar 
as $m_i=17$ (orange curve), $18$ (blue) and $19$ (green).   
The black curve is the result without the magnification bias.
Note that the probability is for the fixed number density of the lens
 $n=1h^3 {\rm Mpc}^{-3}$ and the result scales as $P \propto n$.
}
\label{lensing_prob}
\vspace*{0.5cm}
\end{figure}

In this section, we discuss the lensing probability that a distant quasar
 is lensed by the foreground negative-mass object or the Ellis wormhole.
We assume that the lens objects are uniformly distributed in the
 Universe.
We basically follow the calculation in the SQLS (Oguri et. al. 2012).
The basics of the lensing probability is discussed in
 Schneider et al. (1992, 2006).
The probability is roughly estimated as a product of 
 the distance to the quasar $D_S$, the number density of the lens $n$,
 and the cross section $\sigma$, i.e. $P \sim n \sigma D_S$.
Here, the cross section is defined in the lens plane and is roughly given by
 $\sigma \sim \pi (D_L \theta_{\rm E})^2$.

First, we evaluate the cross section.
The SQLS searched multiple quasar images with the image separation
 range of $1^{\prime \prime} < \Delta \theta < 20^{\prime \prime}$ and
 with the flux ratio $\mu_+/\mu_-$ less than $10^{-0.5}$.
Then, the cross section to form multiple images is
\beqa
  \sigma = 2 \pi D_L^2
  \int_{\beta_{\rm min}}^{\beta_{\rm max}} d\beta \beta
  ~\frac{\Phi(L/\mu,z_s)}{\mu \Phi(L,z_s)} 
  ~\Theta\left(10^{-0.5} - \frac{\mu_+}{\mu_-} \right) \nonumber \\
  \times \Theta(\Delta \theta - 1^{\prime \prime})
  \Theta(20^{\prime \prime} - \Delta \theta)~~~
\label{cross_section}
\eeqa
where $\Theta (x)$ is the step function: $\Theta (x)=1$ for
 $x \ge 0$ and $\Theta (x)=0$ for $x<0$.
$\beta_{\rm min}$ and $\beta_{\rm max}$ are the minimum and maximum source
 positions to form the multiple images:
$(\beta_{\rm min},\beta_{\rm max})$ $=(2\theta_{\rm E},\infty)$,
 and $(0,\infty)$ for the negative-mass lens and the Ellis wormhole.
Note that the image separation $\Delta \theta$ and the flux ratio
 $\mu_+/\mu_-$ in the integrand of Eq.(\ref{cross_section}) are a
 function of $\beta$.
$\Phi(L,z_s)$ is the quasar luminosity function at $z_s$ and 
 this term is so called the magnification bias which is a correction term
 for flux-limited survey.
We use the same luminosity function as that used in Oguri et al. (2012)
 (see e.g. Croom et al. 2009; Richards et al. 2006). 

We assume that the number density of the lens is constant in comoving scale.
Then the lensing probability for the quasar at $z_s$
 with an apparent magnitude $m_i$ is
\beq
  P(z_s,m_i) = \int_0^{z_s} dz_l \frac{cdt}{dz_l} n \left( 1+z_l \right)^3
    \sigma 
\eeq
where $z_l$ is the lens redshift and $cdt/dz_l=1/(H(z_l) (1+z_l))$ 
 with the Hubble expansion rate $H(z_l)$.

Fig.\ref{lensing_prob} shows the lensing probability for the quasar at the
 redshift $z_s$ with the magnitude $m_i$.
The source redshift is set to be $z_s=1.56$ which is the mean redshift
 of the statistical sample of quasars in the SQLS.
The left panel is for the negative-mass compact object, and the right
 panel is for the Ellis wormhole.
The horizontal axis is the absolute value of the lens mass $|M|$ (left panel)
 and the throat radius $a$ (right panel). 
The quasar magnitudes are $m_i=17$ (orange), $18$ (blue) and $19$ (green), and
 the black curve is the result without the magnification bias.
The result is plotted for the fixed number density of the lens
 $n=1h^3 {\rm Mpc}^{-3}$ and the probability scales as $P \propto n$.

In order to evaluate the observational upper bound 
on the number density $n$ of the lenses, 
we use the likelihood function introduced in Kochanek (1993), 
$\ln L \simeq - \sum_{j=1}^{N_{\rm Q}} P(z_{s,j},m_{i,j})$,
where $z_{s,j}$ and $m_{i,j}$ are the redshift and the apparent 
 magnitude of the j-th quasar.
$N_{\rm Q}$ is the total number of statistical samples of quasars,
 $N_{\rm Q}=50836$.
The data of $z_{s,j}$ and $m_{i,j}$ were downloaded from the SQLS
 website\footnote{http://www-utap.phys.s.u-tokyo.ac.jp/\~{}sdss/sqls/}.

\section{Results}

\begin{figure}
\epsscale{1.}
\plotone{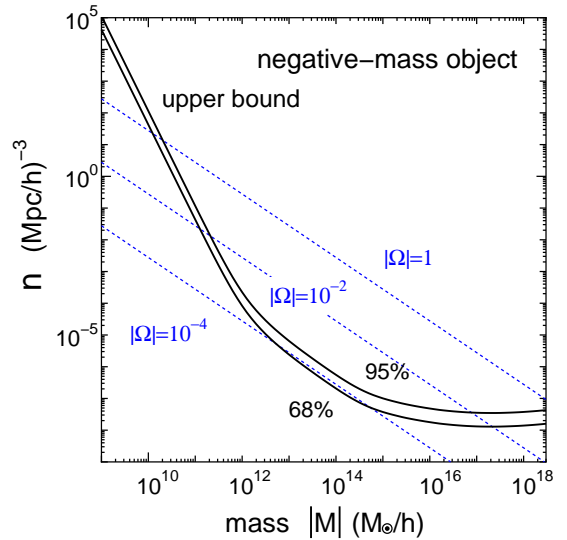}
\caption{
Upper bound on the cosmological number density of 
negative-mass compact objects.
The vertical axis 
denotes the density $n$ ($h^3 {\rm Mpc}^{-3}$). 
The horizontal axis 
denotes 
the absolute value of the mass $|M|$ ($h^{-1} M_{\odot}$). 
The two solid curves correspond to upper bounds of
 $68\%$ and $95\%$ confidence levels.
The blue dashed lines denote the absolute value of the cosmological
 density parameter for the negative mass, $|\Omega|$
 $=10^{-4}, 10^{-2}$ and $1$.
}
\label{upp_bound_dens_negamass}
\vspace*{0.5cm}
\end{figure}

\begin{figure}
\epsscale{1.}
\plotone{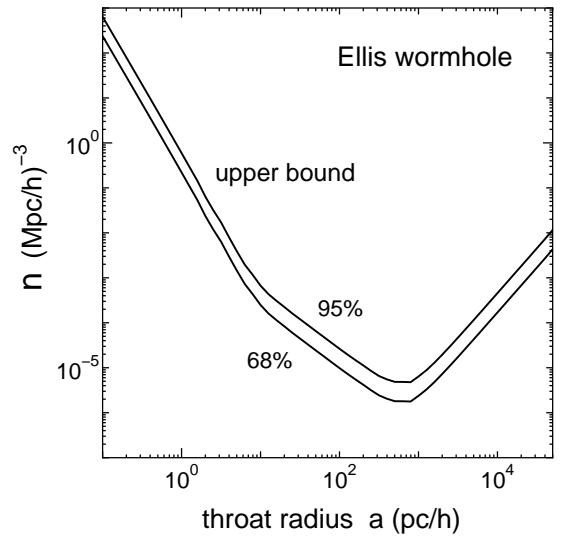}
\caption{
Same as Fig.\ref{upp_bound_dens_negamass}, but for the Ellis wormhole.
The horizontal axis is the throat radius $a$ ($h^{-1}$pc). 
}
\label{upp_bound_dens_ellis}
\vspace*{0.5cm}
\end{figure}

Fig.\ref{upp_bound_dens_negamass} shows the upper bound on the cosmological
 number density of the negative-mass compact objects. 
The vertical axis is the number density $n$ ($h^3 {\rm Mpc}^{-3}$), 
 while the horizontal axis is the absolute value of the mass $|M|$
 $(h^{-1} M_\odot)$. 
The two curves are the upper bound 
at $68\%$ and $95\%$ confidence levels. 
It turns out that the number density is less than
 $n<10^{-7}(10^{-5})h^3 {\rm Mpc}^{-3}$ for $|M|>10^{14}(10^{12})M_\odot$. 
The blue dashed lines show the absolute value of the cosmological density
 parameter, $|\Omega|=10^{-4},10^{-2}$ and $1$.
The density parameter is defined as the mass density divided by the
 cosmological critical density, $\Omega=Mn/\rho_{\rm cr}$.
As shown in the figure, the density parameter is less than
 $|\Omega|<10^{-4}$ for galactic and cluster mass scale
 $|M|=10^{12}-10^{15}M_\odot$.
As a result,
 the negative-mass compact object is less abundant than the galaxy
 with typical luminosity $L^*$ ($n_{\rm gal} \approx 10^{-2}
 {\rm Mpc}^{-3}$) and the galaxy cluster with typical mass $10^{14}M_\odot$
 ($n_{\rm clust} \approx 10^{-4} {\rm Mpc}^{-3}$), which correspond to
 $\Omega_{\rm gal} \approx 0.2$ for the galaxies and $\Omega_{\rm clust}
 \approx 0.3$ for the clusters (e.g. Fukugita \& Peebles 2004)

Fig.\ref{upp_bound_dens_ellis} 
shows the Ellis wormhole cases. 
The horizontal axis is the throat radius $a$ ($h^{-1}$pc).
As shown in the figure, the number density is
 $n<10^{-4}h^3 {\rm Mpc}^{-3}$ for $a=10-10^4$pc. 
As a result, the Ellis wormhole with $a=0.1-10^5$pc is much less 
 abundant than a star ($n_{\rm star} \approx 10^{10} {\rm Mpc}^{-3}$).
Note that our extragalactic constraint is complementary 
to the galactic one by the microlensing that is sensitive 
for the smaller radius $a=100-10^7$km
 (Abe 2010).
The upper-bound curves in Fig.\ref{upp_bound_dens_ellis} 
approach straight lines for very small $a (\ll 0.1 \rm{pc})$ or large
 $a (\gg 10^4 \rm{pc})$.
In the case of a very small $a$, the lens is very close to us since
 $a \propto D_L$ from Eq.(\ref{eins_rad_ellis}) under the fixed
 $\theta_{\rm E}$ and $D_S$, 
while the lens for a very large $a$ is very close to
 the source since $a \propto D_{LS}^{-1/2}$ from Eq.(\ref{eins_rad_ellis}).

\section{Discussion and Conclusion} 

We presented the upper bound on the cosmic abundances of
 the negative-mass compact object and the Ellis wormhole using the
 quasar lens sample in the SQLS data based on SDSS II.
Our main results are summarized in Figs.\ref{upp_bound_dens_negamass}
 and \ref{upp_bound_dens_ellis}.
On-going or future surveys such as
 Pan-Starrs\footnote{http://pan-starrs.ifa.hawaii.edu/public/},
 Dark Energy Survey\footnote{http://www.darkenergysurvey.org/}, 
 Subaru Hyper Suprime-Cam (Miyazaki et al. 2006),
 Large Synoptic Survey Telescope (LSST) \footnote{http://www.lsst.org/lsst/}
 will find much more lensed quasars by the foreground galaxies
 than the SDSS II (Oguri \& Marshall 2010).
Especially, the LSST will find about $8000$ lensed quasars
 which are much more than the 19 lensed quasars in the SQLS.
Therefore, the LSST will provide over 100 times stronger constraint
 than the current upper bound shown in Figs.2 and 3,
 or might lead to a first detection of such exotic lenses.

Note that our constraint for the negative masses can be applied only for
 the compact objects (i.e. its size is smaller than
 the Einstein radius).
As mentioned in the introduction, the negative masses can
 attract each other to form a massive clump.
For such negative-mass clouds, our upper bound in
 Fig.\ref{upp_bound_dens_negamass} becomes weaker.
For instance, one can show that the singular isothermal sphere lens
 composed of the negative masses does not form multiple images.
In this case, the weak lensing is a more powerful tool to identify the
 negative-mass clumps.
The negative masses could act as voids in the universe, 
 since the void is the underdense region compared with the cosmic mean
 density and hence its convergence is {\it negative}.  
Several authors 
reported 
possible detections of voids by
 the weak lensing of background galaxies (e.g. Miyazaki et al. 2002;
 Gavazzi \& Soucail 2007; Shan et al. 2012) and by the integrated
 Sachs-Wolfe effect in the Cosmic Microwave Background
 (e.g. Granett et al. 2008; Ili\'c et al. 2013; Cai et al. 2013).
If the density contrast of the void 
were 
less than $-1$, 
it could be an evidence for the negative masses.  
Because of the low signal-to-noise ratio in the present analyses,  
the next generation surveys 
are awaited.

\acknowledgments
RT thanks Masamune Oguri and Issha Kayo for useful comments about the
 SQLS. 
HA would like to thank Fumio Abe, Matthias Bartelmann, Matt Visser, 
Koji Izumi, Takao Kitamura, and Koki Nakajima 
for stimulating conversations on the exotic lens models. 
RT is supported by Grant-in-Aid for Japan Society for the Promotion of
 Science (No. 25287062).



\end{document}